\documentclass[12pt]{article}

\usepackage{amsmath,amssymb,amsthm,epsf,epsfig}
\usepackage{amsmath,amssymb}

\newcommand{\doublespacing}{\let\CS=\@currsize\renewcommand{\baselinesstrech}
{2.0}\tiny\CS}

\begin{document}

\textwidth 16cm
\newcommand{\bd}{\begin{document}}
\newcommand{\ed}{\end{document}}
\newcommand{\bc}{\begin{center}}
\newcommand{\ec}{\end{center}}
\newcommand{\bfr}{\begin{flushright}}
\newcommand{\efr}{\end{flushright}}
\newcommand{\lt}{\left}
\newcommand{\rt}{\right}
\newcommand{\vs}{\vspace}
\newcommand{\hs}{\hspace}
\newcommand{\beq}{\begin{equation}}
\newcommand{\eeq}{\end{equation}}
\newcommand{\lb}{\linebreak}
\newcommand{\pb}{\pagebreak}
\newcommand{\mb}{\makebox}
\newcommand{\fb}{\framebox}
\newcommand{\mc}{\multicolumn}
\newcommand{\ben}{\begin{enumerate}}
\newcommand{\een}{\end{enumerate}}
\newcommand{\bit}{\begin{itemize}}
\newcommand{\eit}{\end{itemize}}
\newcommand{\ol}{\overline}
\newcommand{\un}{\underline}
\newcommand{\lefq}{\lefteqn}
\newcommand{\ba}{\begin{array}}
\newcommand{\ea}{\end{array}}
\newcommand{\beqa}{\begin{eqnarray}}
\newcommand{\eeqa}{\end{eqnarray}}
\newcommand{\beqas}{\begin{eqnarray*}}
\newcommand{\eeqas}{\end{eqnarray*}}
\newcommand{\bfg}{\begin{figure}}
\newcommand{\efg}{\end{figure}}
\newcommand{\bds}{\begin{displaymath}}
\newcommand{\eds}{\end{displaymath}}
\newcommand{\btb}{\begin{tabbing}}
\newcommand{\etb}{\end{tabbing}}
\newcommand{\para}{\parallel}
\newcommand{\pad}{\partial}
\newcommand{\nn}{\nonumber}
\newcommand{\la}{\leftarrow}
\newcommand{\ra}{\rightarrow}
\newcommand{\lgla}{\longleftarrow}
\newcommand{\lgra}{\longrightarrow}
\newcommand{\La}{\Leftarrow}\newcommand{\Ra}{\Rightarrow}
\newcommand{\Lra}{\Leftrightarrow}
\newcommand{\Lgla}{\Longleftarrow}
\newcommand{\Lgra}{\Longrightarrow}
\newcommand{\bm}{\boldmath}
\newcommand{\lan}{\langle}
\newcommand{\ran}{\rangle}
\renewcommand{\a}{\alpha}
\renewcommand{\b}{\beta}
\newcommand{\g}{\gamma}
\newcommand{\G}{\Gamma}
\renewcommand{\d}{\delta}
\newcommand{\eps}{\epsilon}
\newcommand{\Th}{\Theta}
\newcommand{\s}{\sigma}
\newcommand{\lam}{\lambda}
\newcommand{\D}{\Delta}
\newcommand{\vare}{\varepsilon}
\newcommand{\pr}{\prime}
\newcommand{\ro}{\rho}
\newcommand{\nab}{\nabla}
\newcommand{\m}{\mu}
\newcommand{\Sg}{\Sigma}
\newcommand{\p}{\pi}
\newcommand{\R}{I\!\!R}
\newcommand{\om}{\omega}
\newcommand{\Om}{\Omega}
\newcommand{\ze}{\zeta}
\newcommand{\vart}{\vartheta}
\newcommand{\tri}{\triangle}
\newcommand{\f}{\frac}
\newcommand{\iny}{\infty}
\newcommand{\pro}{\propto}
{\bf{To be appeared in Mod. Phys. Lett. A}}
 \bc {\Large {\bf Coherent
state of the effective mass\vspace{.2cm} harmonic oscillator}} \ec

\vs{.5cm} \bc
{\bf \it Atreyee Biswas{\footnote {e-mail : atreyee11@gmail.com}}\\
Department of Natural Science and Humanities\\
West Bengal University of Technology\\
Salt Lake City, Kolkata-700064}\\

\vspace*{1cm}

{\bf \it Barnana Roy{\footnote{e-mail : barnana@isical.ac.in}}\\
Physics \& Applied Mathematics Unit \\
Indian Statistical Institute \\
Kolkata - 700 108, India.} \ec

\vs{1.5cm}

\bc {\large {\un{Abstract}}} \ec We construct coherent state of the
effective mass harmonic oscillator and examine some of it's
properties. In particular closed form expressions of coherent states for
different choices of the mass function are obtained and it is shown
that such states are not in general $x-p$ uncertainty states. We also compute the associated Wigner functions.\\
\pb

\section{Introduction} Schr\"odinger equation with a position dependent or effective
mass (EMSE) has found applications in the fields of material science and condensed matter physics, such as semiconductors [1], quantum wells and
quantum dots [2,3], 3He clusters [4], graded alloys and semiconductor heterostructures [5-13] etc. It has also been found that such equations appear in very different areas. For example, it has been shown that constant mass Schr\"odinger equations in curved space and those based on deformed commutation relations can be interpreted in terms of EMSE [14-15]. The position dependent effective mass also appear in non-linear oscillator [16] and $\cal {PT}$ symmetric cubic anharmonic oscillator [17].
This has generated a lot of interest in this field and
during the past few years various theoretical aspects of EMSE e.g.
exact solvability [18-24], shape invariance property [25,26], quasi-exact solvability [27], connection to higher dimensional systems [28], supersymmetric or intertwining formulation [29-32], Lie algebraic approach [33,34], Green's functions [35] etc. have been studied widely. Also the effect of space dependent mass on the revival phenomena [36] and time evolution of wave packets [37] have also been studied. But so far our knowledge goes, the coherent states [38] of an EMSE has not been discussed in the literature. Motivated by the fact that the coherent states for the constant mass Schr\"odinger equation have revealed a surprisingly rich structure, in this note we shall construct coherent state of an EMSE
with harmonic oscillator spectrum by utilising supersymmetric quantum mechanics [39-41] based raising and lowering operators. In this context it should be mentioned that supersymmetric quantum mechanics based raising and lowering operators have found significant application in the construction of coherent states of constant mass Schr\"odinger equation for various potentials [42-44]. We shall examine different properties
of the EMSE coherent states. In particular we shall examine the behaviour of such a state with respect to the physical $x-p$ uncertainty relation. The possibility of squeezing and the effect of
variation of mass on it will also be examined. Finally we shall compute the Wigner function and show that it takes negative values in certain ranges of the parameters.

\section{Effective mass harmonic oscillator} It may be
noted that in the case of an effective mass the kinetic energy (and
consequently the Hamiltonian) can be defined in several ways. The
most general Hamiltonian can be wriiten in the form [45]
\begin{equation}
H=\frac{1}{4}\left(m^{\alpha}(x)pm^{\beta}(x)pm^{\gamma}(x)+m^{\gamma}(x)pm^{\beta}(x)pm^{\alpha}(x)\right)+V(x)
\end{equation}
where $\alpha$, $\beta$ and $\gamma$ are parameters satisfying the
constraint $\alpha+\beta+\gamma=-1$. Clearly there are different
Hamiltonians depending on the choices of the parameters. Here we
shall work with the BenDaniel-Duke form which corresponds to the
choice $\alpha=\gamma=0,\beta=-1$ [46]. In this case the
Hamiltonian is invariant under instantaneous Galilean transformation
[47]. The corresponding Schr\"{o}dinger equation is given by
\begin{equation}
-\frac{d}{dx}\left(\frac{1}{2m(x)}\frac{d\psi(x)}{dx}\right) +
V(x)\psi(x) = E\psi(x) \label{sch1}
\end{equation}
The above equation can be solved in many ways e.g, supersymmetric methods [29-32]
,  using point
canonical transformations [48-50] etc. Here we shall follow the former method and consider
the operators
\begin{eqnarray}
A\psi &=& \frac{1}{\sqrt{2m}}\frac{d\psi}{dx} + W\psi\\ \nonumber
\label{a} A^{\dagger}\psi &=&
-\frac{d}{dx}\left(\frac{\psi}{\sqrt{2m}}\right) + W\psi\\
\label{adag}
\end{eqnarray}
where the function $W(x)$ is known as the superpotential. Then the
Hamiltonians
\begin{equation}
H = A^{\dagger}A = -\frac{1}{2m}\frac{d^2}{dx^2} -
\left(\frac{1}{2m}\right)^{'}\frac{d}{dx} -
\left(\frac{W}{\sqrt{2m}}\right)^{'} + W^2 \label{h1}
\end{equation}
\begin{equation}
{\tilde H} = AA^{\dagger} = -\frac{1}{2m}\frac{d^2}{dx^2} -
\left(\frac{1}{2m}\right)^{'}\frac{d}{dx} -
\left(\frac{W}{\sqrt{2m}}\right)^{'} + W^2 +
\frac{2W^{'}}{\sqrt{2m}} -
\left(\frac{1}{\sqrt{2m}}\right)\left(\frac{1}{\sqrt{2m}}\right)^{''}
\label{h2}
\end{equation}
are isospectral and the corresponding potentials $V$ and $\tilde V$ are
given by
\begin{eqnarray}
V &=& -\left(\frac{W}{\sqrt{2m}}\right)^{'} + W^2 \\ \nonumber
\label{pot1} {\tilde V} &=& -\left(\frac{W}{\sqrt{2m}}\right)^{'} + W^2 +
\frac{2W^{'}}{\sqrt{2m}} -
\left(\frac{1}{\sqrt{2m}}\right)\left(\frac{1}{\sqrt{2m}}\right)^{''}\\
\label{pot2}
\end{eqnarray}

Then from (\ref{h1}) and (\ref{h2}) it follows that
\begin{equation}
[A , A^{\dagger}] = \frac{2W^{'}}{\sqrt{2m}} -
\left(\frac{1}{\sqrt{2m}}\right)\left(\frac{1}{\sqrt{2m}}\right)^{''}
\label{cond}
\end{equation}
Clearly if $A$ and $A^{\dagger}$ are to be interpreted in the same
way as the standard harmonic oscillator annihilation and creation
operator respectively then we should have $[A,A^{\dagger}]=1$ and in
this case we obtain from (\ref{cond})
\begin{equation}
2W(x) = \left(\frac{1}{\sqrt{2m}}\right)^{'} + \int^x \sqrt{2m(z)}
dz \label{W}
\end{equation}
Thus for a given mass $m(x)$ the superpotential $W(x)$ can be
determined from the relation (\ref{W}). Thus for such
superpotentials $A^{\dagger}$ and $A$ have properties same as the
standard bosonic creation and annihilation operators respectively (for example, $A^\dagger \psi_n = \sqrt{n+1}\psi_{n+1}~,~A\psi_n = \sqrt{n}\psi_{n-1}$).
In this case the spectrum of $H=A^{\dagger}A$ and ${\tilde
H}=AA^{\dagger}$ are esentially that of the constant mass harmonic
oscillator. In particular the spectrum and eigenfunctions of $H$ are
given by \beq E_n = \left(n+\f{1}{2}\right)~~,~~ \psi_n(x) =
\f{1}{\sqrt{\sqrt{2\pi}2^nn!}}~ [2m(x)]^{1/4}~e^{-{\bar
x}^2/4}H_n({\bar x}/{\sqrt 2})\label{wf}\eeq where \beq {\bar x} =
\int^x \sqrt{2m(y)}~dy\eeq

\section{Coherent state and it's properties}
There are a number of ways to construct coherent states. However, in
view of the fact that the operators $A$ and $A^\dagger$ satisfy the
relation $[A,A^\dagger]=1$, the coherent state can be constructed
using the displacement operator technique. Thus we define the
unitary displacement operator $D(z)$ as \beq D(z) =
e^{(zA^\dagger-z^*A)}\eeq and the coherent state is given by \beq
\left|\psi\right>_{cs} = e^{(zA^\dagger-z^*A)}\left|0\right> =
e^{-\f{1}{2}|z|^2}\sum_{n=0}^\infty
\f{z^n}{n!}\left|n\right>\label{cs1}\eeq Now using (\ref{wf}) and
the properties of Hermite polynomials, the coherent state in
coordinate representation is found to be \footnote[1]{The coherent state (\ref{cs2}) can also be obtained as an eigenstate of the annihilation operator $A$.}\beq \psi_{cs}(x) =
\f{1}{\sqrt{\sqrt{2\pi}}}e^{-(|z|^2/2 - z^2/2)}[2m(x)]^{\f{1}{4}}e^{-({\bar
x}-2z)^2/4}\label{cs2}\eeq
In this context it must be mentioned that this formalism of constructing coherent states works only when Eqns.(7) and (10) holds, i.e. the formalism is not suitable for constructing coherent state for a general potential. More specifically, the coherent states can be constructed using the present formalism only when the system has Heisenberg-Weyl, SU(2) or SU(1,1) symmetry. For systems whose symmetry structure is different from these mentioned before, a convenient way to construct coherent state is to follow the Gazeau-Klauder formalism [52,53] which requires the knowledge of the spectrum and the eigenfunctions.

It is not difficult to show that (\ref{cs2}) shares many of the properties of standard coherent states. For example, the time dependent coherent state is given by \beq \psi_{cs}(x,t) = e^{-iHt}\psi_{cs}(x) = \displaystyle{e^{-(|z|^2-z'^2+it)/2}} \sqrt{2m(x)}~\displaystyle{e^{-(\bar x - 2z')^2/4}}\label{cst}\eeq where $z'
= ze^{-it}$. Thus the coherent state at $t=0$ remains a coherent state at $t\neq 0$ with a different parameter $z'=ze^{-it}$.

We shall now examine the possibility of squeezing. To this end we consider two Hermitian operators \beq X = \f{A+A^\dagger}{\sqrt{2}}~~,~~ Y =
\f{-i(A-A^\dagger)}{\sqrt{2}}\eeq where $\Delta X = <X^2> - <X>^2$. Then using the properties of $A^\dagger, A$ it can be easily shown that \beq (\Delta X)
= \f{1}{2}~~,~~(\Delta Y) = \f{1}{2}\eeq  Thus the coherent state
(\ref{cs2}) saturates the uncertainty relation \beq (\Delta
X)(\Delta Y) \geq \f{1}{4}\label{xy}\eeq and is a minimum
uncertainty state with respect to the relation (\ref{xy}). However,
the operators $X$ and $Y$ are not physical position ($x$) or momentum ($p$)
operators. Thus it would be interesting to examine the behaviour
of (\ref{cs2}) with respect to the physical $x-p$ uncertainty relation \beq (\Delta
x)(\Delta p) \geq \f{1}{4}\label{uncer}\eeq  We note that if $\Delta x$ and $\Delta p$ are each greater than $1/2$, then the above inequality is always true. However, the above inequality holds even if one of $\Delta x$ or $\Delta p$ is less than $1/2$ and the other sufficiently large. In this case the state is squeezed. To get a quantitative measure of squeezing we introduce squeezing parameters $S_x$ and $S_p$ defined by :\beq
S_x = 2(\Delta x) -1~~,~~S_p = 2(\Delta p) -1\eeq Thus the state would be squeezed if either $S_x<0$ or $S_p<0$.

We shall now study how far the space dependent mass $m(x)$ influences various features of the coherent state.  \vspace{.2cm}

{\bf{\underline {Case 1}.}} Let us consider the following mass profile which is considered by some authors in graded alloys [51]
\beq 2m(x) = cosh^2(\alpha x),~~-\infty<x<\infty\label{m1}\eeq so that for
$\alpha = 0$ we recover the constant mass harmonic
oscillator. In this case we have
\beq
\psi_n(x) = {\displaystyle
\f{1}{\sqrt{\sqrt{2\pi}2^nn!}}}~ \sqrt{cosh(\alpha x)}~e^{-{\bar
x}^2/4}H_n({\bar x}/{\sqrt 2})
\eeq
and
\beq
\bar x = {\displaystyle \f{sinh(\alpha x)}{\alpha}},~~-\infty<{\bar x}<\infty
\eeq
and the effective mass coherent state is found to be
\beq
\psi_{cs}(x) =
\f{1}{\sqrt{\sqrt{2\pi}}}~e^{-(|z|^2/2 - z^2/2)}{\sqrt{cosh(\alpha x)}}~e^{-(sinh(\alpha x)/\alpha-2z)^2/4}\label{cs3}\eeq
\vspace{.2cm}

Next we shall examine a very important aspect of the coherent states, namely their behaviour with respect to the physical $x-p$ uncertainty relation. To this end we have evaluated the uncertainty product $(\Delta x)(\Delta p)$ and the
squeezing parameters $S_x,S_p$ for the state (\ref{cs3}) for
different values of the mass parameter $\alpha$. From Fig. $1a)$ we
find that for larger values of $\alpha$, the uncertainty product is larger for smaller values of $z$ but for larger values of $z$, it stabilises and is nearly equal for both values of $\alpha$. However it always remains greater than $0.25$ and consequently the inequality (\ref{uncer}) holds.
\vspace{.2cm}

From Fig. $2(a)$ it is seen that $S_x$ is
negative for some values of the parameters. In particular for fixed $\alpha$, squeezing increases for larger $z$. Thus the coherent state exhibits squeezing in the $x$ quadrature.
We have plotted $S_p$ in Fig. $3(a)$. From Fig. $3(a)$ we find that for larger $\alpha$, $S_p$ is smaller for smaller $z$. Subsequently it increases as $z$ increases. However for all values of the parameters $S_p>0$
implying absence of squeezing in the $p$ quadrature. The same pattern
can also be observed for other values of $\alpha$ and $z$. We note that this non classical
behaviour is quite different from the standard harmonic oscillator
coherent states (which are minimum $x-p$ uncertainty states and never
shows squeezing).
\vspace{.2cm}

We shall now compute the Wigner function for the coherent state (\ref{cs3}). The Wigner function is defined as
\beq
W(x,p) = \f{1}{\pi}\int_{-\infty}^\infty \psi^*_{cs}(x-y)e^{2ipy}\psi_{cs}(x+y)~dy\label{wigner}\eeq
We recall that the Wigner function may or may not take negative values. However, negativity of the Wigner function is a sufficient condition for the state to be nonclassical. In Fig. $4(a)$ we have plotted the Wigner function $W(x,p)$ against $\alpha$ and $z$. It can be seen from the Fig. $4(a)$ that the Wigner function does take negative values and it confirms non classical nature of the coherent state (\ref{cs3}).
\vspace{.25cm}

{\bf{\underline {Case 2}.}} We now consider another mass profile given by the following which is found to be useful for studying transport properties in semiconductors [29-32,54]
given by \beq 2m(x) = \left(\f{\alpha + x^2}{1 +
x^2}\right)^2,~~-\infty<x<\infty\eeq In this case
\beq
{\bar x} = x + (\alpha-1)\arctan x,~~-\infty<{\bar x}<\infty
\eeq
The corresponding coherent state is given by
\beq
\psi_{cs}(x) =
\f{1}{\sqrt{\sqrt{2\pi}}}~e^{-(|z|^2 - z^2)/2}{\sqrt{\f{\alpha+x^2}{1+x^2}}}~e^{-(x+(\alpha-1)\arctan  x-2z)^2/4}\label{cs4}\eeq

It may be noted that in this case the mass distribution has different shapes in the ranges $\alpha <1$ and $\alpha>1$ (Fig. $5$). Now as in the last case we have evaluated the uncertainty product and from Fig. $1(b)$ we find that the inequality (\ref{uncer}) holds. So the coherent state (\ref{cs4}) is not a $x-p$ minimum uncertainty state. However, for fixed $\alpha$, the uncertainty product gets smaller as $z$ increases. So for very large $z$ it behaves like a minimum uncertainty state.
\vspace{.2cm}

We now compute the squeezing parameters. From Fig. $2(b)$ and Fig. $3(b)$ we find that $S_x$ and $S_p$ behave differently for $\alpha>1$ and $\alpha<1$. For $\alpha>1$, $S_x$ starts from a relatively small positive value and $S_p$ starts from a small negative value. As $z$ increases both $S_x$ and $S_p$ stabilize while retaining their character. For $\alpha<1$ the scenario is exactly opposite. Interestingly in both the ranges of $\alpha$, $S_x$ remains positive while $S_p$ is always negative. So in this case the coherent state exhibits squeezing in the $p$ quadrature but not in the $x$ quadrature.
\vspace{.2cm}

We now compute the Wigner function using (\ref{wigner}) with $\psi_{cs}(x)$ given by (\ref{cs4}). It can be seen from Fig $4(b)$ that the Wigner function does take a small negative value indicating the non classical nature of the state (\ref{cs4}).

\section{Conclusion} Here we have constructed coherent state of effective mass harmonic oscillator with two different mass distributions. It has been shown that in both the cases the coherent states exhibit squeezing and the non classical nature of these states is confirmed by the negativity of the corresponding Wigner functions. In the second case, since the mass function is an increasing function of $\alpha$, squeezing can be increased using larger $\alpha$.
Let us note that in the case of EMSE coherent state, the
displacement operator coherent state and the annihilation operator
coherent state are the same as in the case of coherent states of
harmonic oscillator in constant mass Schr\"{o}dinger equation.
But, unlike the constant mass case, the EMSE coherent state is not
a $x-p$ uncertainty state. It exhibits squeezing. The effect of
variation of mass function on the coherent state properties are
evident from the figures (1) - (4). Considering the fact that
coherent states for systems other than the harmonic oscillator in
constant mass Schr\"{o}dinger equation have attracted much
attention for several years [55-71], it would be interesting to
construct coherent states for non harmonic type effective mass
systems. In such cases it would be useful to carry out the
construction using, for example, the Gazeau-Klauder formalism
[52,53].\\

\newpage

{\large{\bf{References}}\\
\begin{enumerate}
\item G. Bastard, Wave Mechanics Applied to Semiconductor
Heterostructures (Les Editions de Physique,Les Ulis,France,1988).
\item L.I. Serra and E. Lipparini, Europhys.Lett {\bf 40}, 667
(1997). \item P. Harrison, Quantum Wells, Wires and Dots (John
Wiley and Sons, 2000).
 \item M.Barranco, M.Pi, S.M.Gatica,
E.S.Hernandez and J.Navarro, Phys.Rev.B {\bf 56} 8997 (1997).
\item T. Gora and F. Williams, Phys. Rev. {\bf 177} 1179 (1969).
\item R.A.Marrow, Phys.Rev.B {\bf 27} 2294 (1985). \item
R.A.Morrow, Phys.Rev.B {\bf 36} 4836 (1987). \item W.
Trzeciakowski, Phys.Rev.B {\bf 38} 4836 (1987). \item I. Galbraith
and G.  Duggan, Phys.Rev.B {\bf 38} 10057 (1988). \item K. Young,
Phys.Rev.B {\bf 39} 13434 (1989). \item G.T. Einevoll, P.C. Hemmer
and J. Thomsen, Phys.Rev.B {\bf 42} 3485 (1990). \item G.T.
Einevoll, Phys.Rev.B {\bf 42} 3497 (1990). \item C.Weisbuch and B.
Vinter, Quantum Semiconductor Heterostructures (Academic Press,
New York,1993).

\item C. Quesne and V.M. Tkachuk, J.Phys.A {\bf 37}, 4267 (2004) .
\item B. Bagchi, A.Banerjee, C. Quesne and V.M. Tkachuk, J.Phys.A
{\bf 38}, 2929 (2005) . \item P.M.Mathews and M.Lakshmanan,
Quart.Appl.Math. {\bf 32} 215 (1974). \item A.Mostafazadeh,
J.Phys.A {\bf 40} 6557 (2005); Erratum ibid, {\bf 38} 8185 (2005).
\item L. Dekar et al, J.Math.Phys. {\bf 39} 2551 (1998). \item L.
Dekar et al, Phys.Rev.A {\bf 59} 107 (1999). \item B. Bagchi et
al, Czech.J.Phys. {\bf 54} 1019 (2004). \item B.Bagchi et al,
Mod.Phys.Letts.A {\bf 19} 2765 (2004). \item J. Yu, S.H. Dong and
G.H. Sun, Phys.Letts.A {\bf 322} 290 (2004). \item J. Yu and S.H.
Dong, Phys.Letts.A {\bf 325} 194 (2004). \item A.Ganguly, S.Kuru,
J.Negro, L.M.Nieto, Phys.Letts.A {\bf 360} 228 (2006). \item K.
Samani and F. Loran, e-print arXive: quant-ph/0302191. \item B.
Bagchi, A. Banerjee, C. Quesne and V.M. Tkachuk, J.Phys.A {\bf 38}
2929 (2005). \item R. Koc, E. K\"orc\"uk and M. Koca, J.Phys.A
{\bf 35} L1 (2002). \item C. Quesne and V.M. Tkachuk, J.Phys.A
{\bf 37} 4267 (2004). \item A.R. Plastino, A. Rigo, M. Casas, F.
Gracias and A. Plastino, Phys.Rev A {\bf 60}, 4398 (1999) . \item
B. G\"onul, B. G\"onul, D. Tutcu and O. \"Ozer, Mod.Phys.Letts.A
{\bf 17}, 2057 (2002) . \item C. Quesne, Ann.Phys {\bf 321}, 1221
(2006) . \item A.Ganguly and L.M.Nieto, J.Phys.A {\bf 40} 7265
(2007). \item B. Roy and P. Roy, J.Phys.A {\bf 35}, 3961 (2002).
\item R. Koc and M. Koca, J.Phys.A {\bf 36}, 8105 (2003). \item L.
Chetouani, L. Dekar and T.F. Hamman, Phys.Rev.A {\bf 52} 82
(1995). \item A.G. Schmidt, Phys.Letts.A {\bf 353}, 459 (2006).
\item A.G. Schmidt, Phys.Scr {\bf 75}, 480 (2007).
\item J.R.Klauder and B.S.Skagerstam, Coherent States-Applications in Physics and Mathematical Physics, World Scientific,Singapore (1985)\\
\item G.Junker, Supersymmetric Methods in Quantum and Statistical
Physics (Springer, 1996). \item F.Cooper, A.Khare and U.Sukhatme,
Phys. Rep. {\bf 25} 268 (1995) . \item B.K.Bagchi, Supersymmetry
in Quantum and Classical Mechanics (Chapman and Hall/CRC, 2001).
\item M.G.Benedict and B.Molnar, Phys.Rev.A {\bf 60} R1737 (1999).
\item A.H.Kinani and M.Daoud, Phys.Letts.A {\bf 283} 291 (2001).
\item J.P.Antoine et al, J.Math.Phys. {\bf 42} 2349 (1999). \item
O. Von Roos, Phys.Rev.B {\bf 27} 7547 (1983). \item D.J. BenDaniel
and C.B. Duke, Phys.Rev {\bf 152}, 683 (1966). \item
J.-M.L\'{e}vy-Leblond, Phys.Rev A {\bf 52}, 1845 (1995). \item B.
Roy and P. Roy, preprint, quant-ph /0106028. \item A.D. Alhaidary,
Phys.Rev.A {\bf 66}, 042116 (2002). \item B. G\"onul, O. \"Ozer,
B. G\"on\"ul and F. \"Uzg\"un, Mod.Phys.Letts.A {\bf 17} 2453
(2002). \item V.Milanovic and Z.Ikonic, Phys.Rev.B {\bf 54} 1998
(1996). \item J.R. Klauder, J.Phys.A {\bf 29}, L293 (1996). \item
J.-P. Gazeau and J.R. Klauder, J.Phys.A {\bf 32}, 123(1999). \item
R.Koc et al, Eur.Phys.Jour.B {\bf 48} 583 (2005). \item M.M.Nieto
and L.M.Simmons,Jr., Phys.Rev.D {\bf 20} 1332, 1342 (1979). \item
M.G.A.Crawford and E.R.Vrscay, Phys.Rev.A {\bf 57} 106 (1998).
\item C.C.Gerry and J.Keifer, Phys.Rev.A {\bf 37} 665 (1988).
\item J.R.Klauder, J.Phys.A {\bf 29} L293 (1996). \item P.Majumdar
and H.S.Sharatchandra, Phys.Rev.A {\bf 56} R3322 (1997). \item
R.F.Fox, Phys.Rev.A {\bf 59} 3241 (1999). \item M.G.A.Crawford,
Phys.Rev.A {\bf 62} 12104 (2000). \item S.A.Polshin, J.Phys.A {\bf
33} L357 (2000). \item N.Gurappa et al, Phys.Rev.A {\bf 61} 34703
(2000). \item M.M.Nieto, Mod.Phys.Letts.A {\bf 16} 2305 (2001).
\item B.Roy and P.Roy, Phys.Letts.A {\bf 296} 187 (2002). \item
A.Chenaghlou and H.Fakhri, Mod.Phys.Letts.A {\bf 17} 1701 (2002);
ibid {\bf 19} 2619 (2004). \item H.Fakhri and A.Chenaghlou,
Phys.Letts.A {\bf 310} 1 (2003). \item A.Chenaghlou and O.Faizy,
J.Math.Phys. {\bf 48} 112106 (2007); ibid {\bf 49} 022104 (2008).
\item T.Fukui and N.aizawa, Phys.Letts.A {\bf 180} 308 (1993).
\item A.B.Balantekin et al, J.Phys.A {\bf 32} 2785 (1999); ibid
{\bf 35} 9063 (2002). \item A.Jellal, Mod.Phys.Letts.A {\bf 17}
671 (2002).

\end{enumerate}

\begin{center}
\begin{figure}
\includegraphics[height=4in]{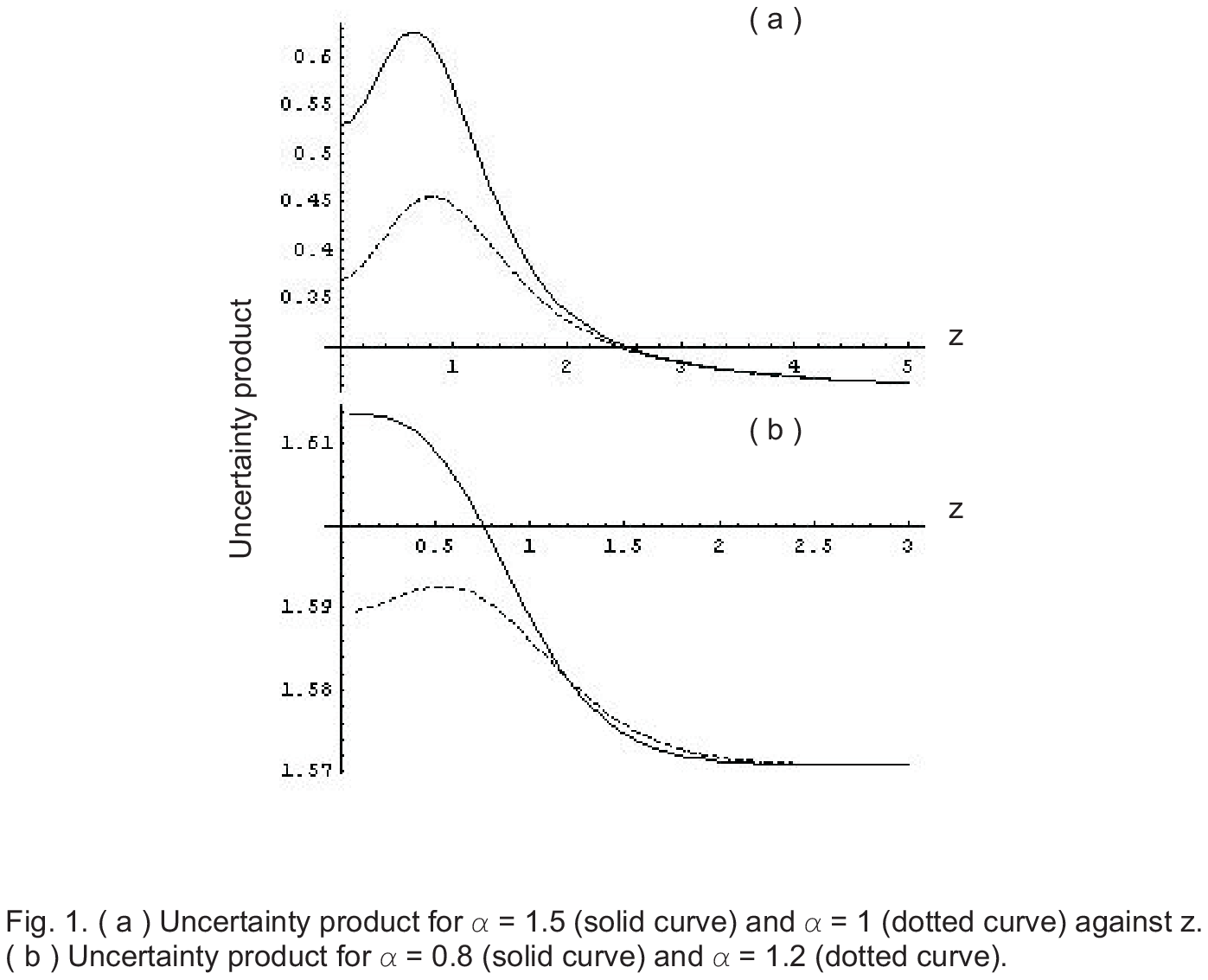}\\
\includegraphics[height=4in]{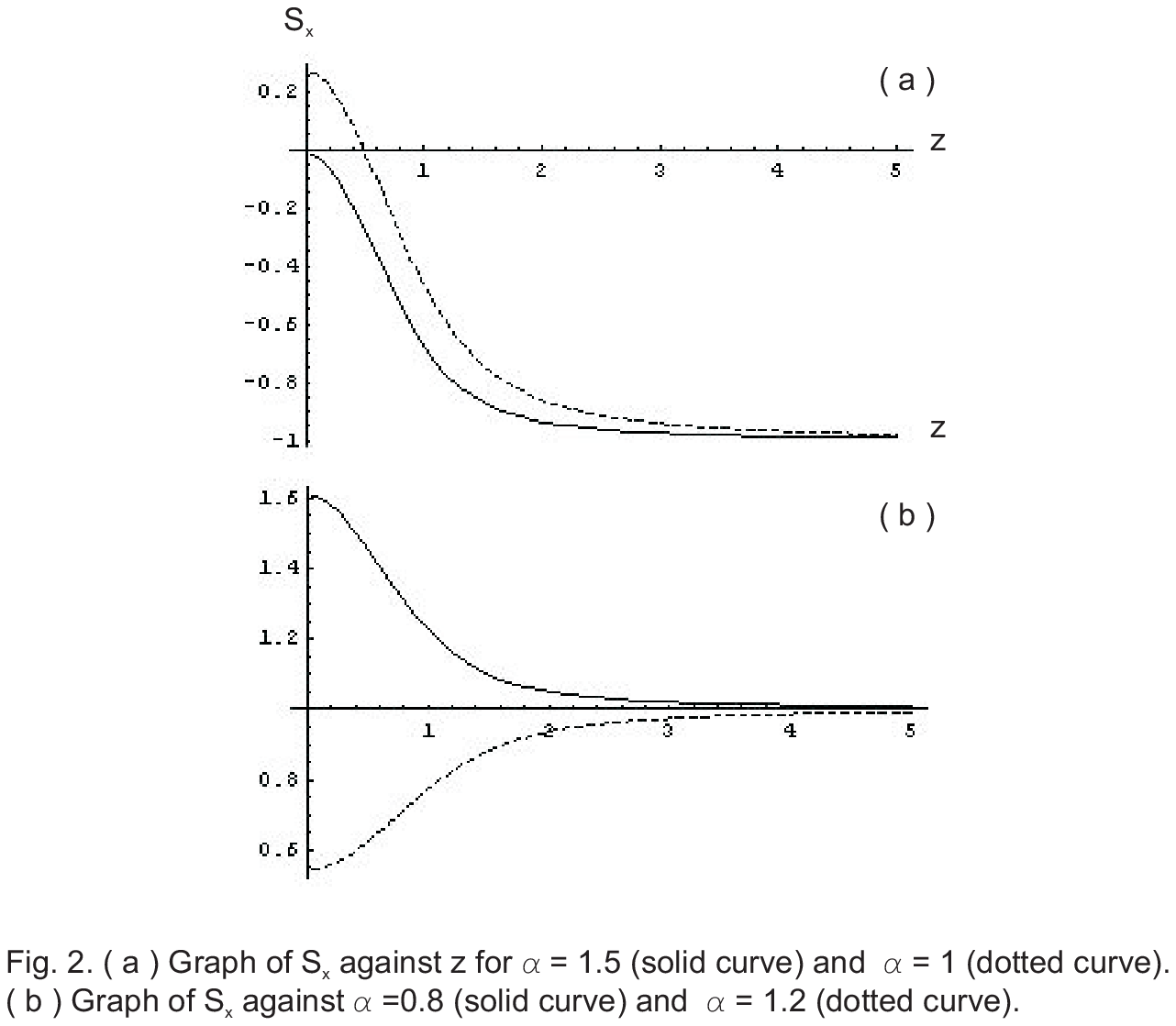}
\end{figure}
\end{center}

\begin{center}
\begin{figure}
\includegraphics[height=4in]{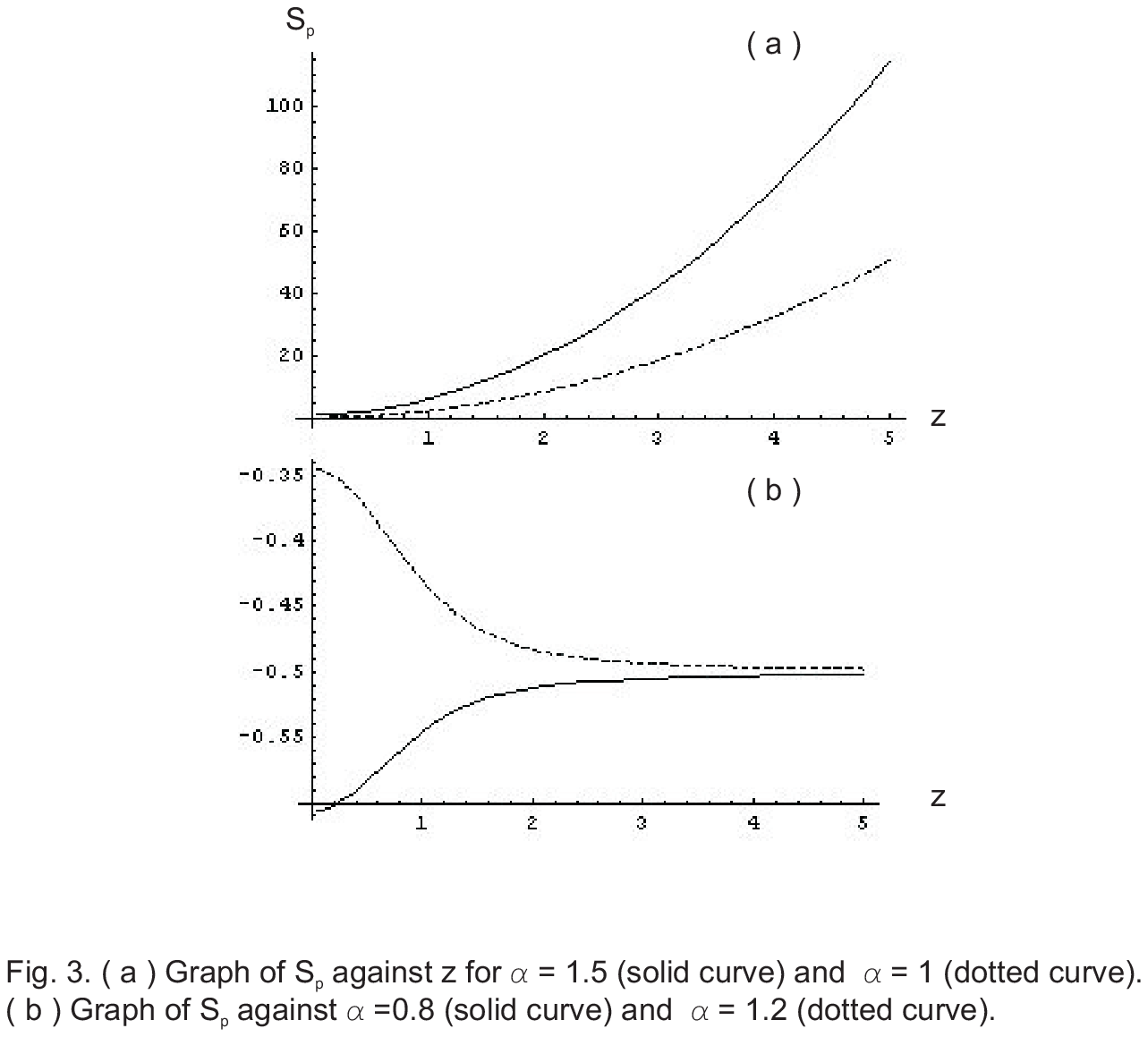}\\
\includegraphics[height=4in]{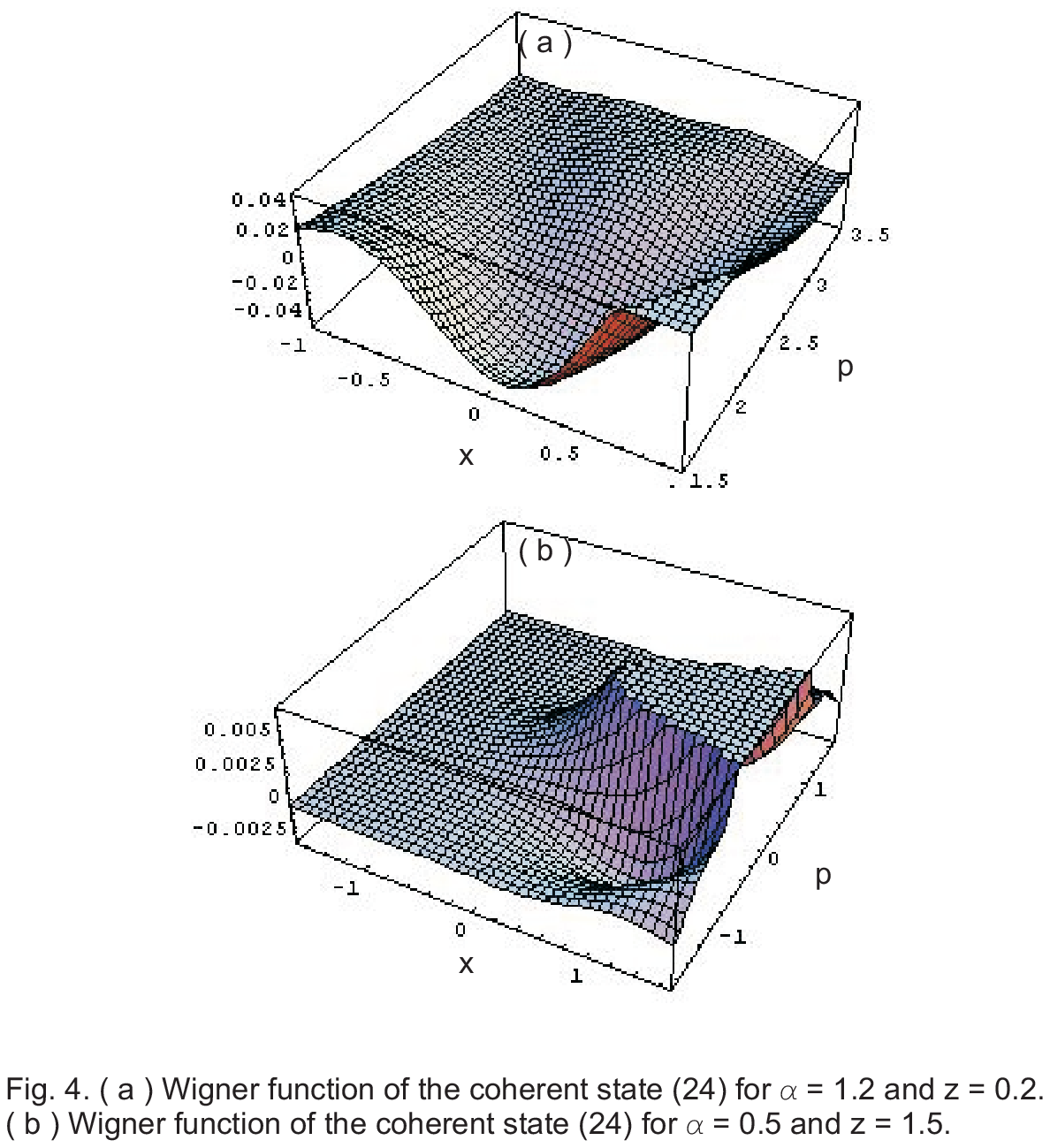}
\end{figure}
\end{center}

\begin{center}
\begin{figure}
\includegraphics[height=3.25in]{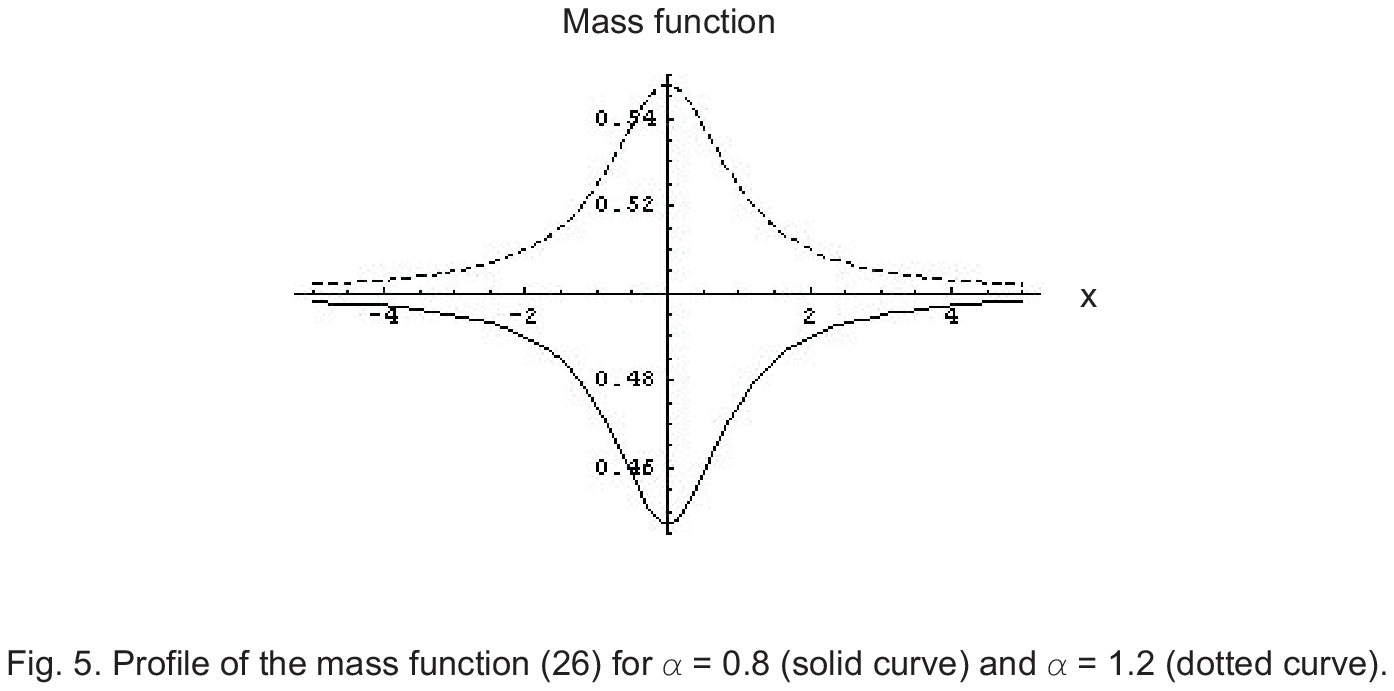}
\end{figure}
\end{center}

 \ed